\documentclass[12pt,epsf]{article}

\usepackage{epsfig,latexsym,amsfonts,amsmath,amsthm,amssymb,amsbsy,multirow,slashed,color,mathrsfs,subfigure,wrapfig}
\usepackage[font={footnotesize,sl},bf]{caption}

\usepackage{paralist}

\newcommand\be{\begin{equation}}
\newcommand\bea{\begin{eqnarray}}
\newcommand\ee{\end{equation}}
\newcommand\eea{\end{eqnarray}}

\newcommand{\bdm}{\begin{displaymath}}
\newcommand{\edm}{\end{displaymath}}

\newcommand{\f}[2]{\frac{#1}{#2}}

\usepackage{color}

\begin{document}
\begin{titlepage}
\hfill

\vspace*{20mm}
\begin{center}
{\Large \bf Black holes vs. firewalls and thermo-field dynamics}

\vspace*{15mm}
\vspace*{1mm}

Borun D. Chowdhury

\vspace*{1cm}

{Institute for Theoretical Physics, University of Amsterdam,\\
Science Park 904, Postbus 94485, 1090 GL Amsterdam, The Netherlands \\
{\it b.d.chowdhury@uva.nl}
}

\vspace*{1cm}
\end{center}

\begin{abstract}
In this essay, we examine the implications of the ongoing {\em black holes vs. firewalls} debate for the thermo-field dynamics of black holes by analyzing a CFT in a thermal state in the context of AdS/CFT. We argue that the themo-field doubled copy of the thermal CFT should be thought of not as a fictitious system, but as the image of the CFT in the heat-bath. While this idea was proposed earlier by Papadodimas et al., our following conclusions differ from theirs. In case of strong coupling between the CFT and the heat-bath this image allows for free infall through the horizon and the system is described by a black hole. Conversely, firewalls are the appropriate dual description in case of weak interaction of the CFT with its heat bath.
\end{abstract}

\vskip 2cm

\begin{center}
Essay written for the Gravity Research Foundation \\ 2013 Awards for Essays on Gravitation.
\vskip 1cm

March 31, 2013
\end{center}
\end{titlepage}

\section*{Introduction}

The recent debate on firewalls~\cite{Almheiri:2012rt} has brought the information paradox and the nature of black hole horizons into sharp focus.  Building on Mathur's information theoretic argument in~\cite{Mathur:2009hf}, which demonstrates the robustness of information paradox, Almheiri et al. showed that the basic axioms of black hole complementarity~\cite{Susskind:1993if} are mutually inconsistent. Considering the validity of effective field theory outside the stretched horizon sacrosanct, they opined that after the black hole has evaporated away half its entropy, an infalling observer perceives the horizon as a {\em firewall} and cannot fall in unharmed.

AdS/CFT, an equivalence between asymptotically anti-de Sitter (AdS) spacetimes and conformal field theories (CFT), is a promising candidate for a  non-perturbative description of quantum gravity. Thus, it is highly desirable to investigate the firewall phenomenon in this context. It was shown in~\cite{Avery:2013exa} that the dual to a black hole which has evaporated half its entropy is a thermal state in the CFT. One can prepare such a state by coupling the CFT to a heat bath. It was argued in~\cite{Avery:2013exa} that such a thermal CFT is dual to a firewall if the heat bath plays no role in the subsequent evolution of an excitation on the CFT. However, note that in~\cite{Maldacena:2001kr} Maldacena had argued that the dual to an eternal AdS black hole involves two copies of CFT in a thermo-field vacuum state. Since, in such a state each copy of the CFT by itself is described by a thermal density matrix, at least in some cases the dual to a thermal CFT is a black hole instead of a firewall.

In this essay we elaborate on this idea and investigate what black holes and firewalls teach us about thermalization in the dual field theory. We  argue that while in  thermo-field dynamics~\cite{doi:10.1142/S0217979296000817}, the doubled copy of a thermal state is understood to be fictitious, it should be thought of as the {\em image} of a system in its heat bath.  We further argue that there is a smooth firewall free horizon in the geometric description only when the dual CFT and its thermo-field double interact, albeit indirectly, through the heat bath. Conversely, in the cases when they do not interact the dual description has a firewall. Thus, while entanglement is necessary for smooth horizons,  it is not sufficient.\footnote{The idea that  the doubled copy of a thermal state should be thought of as the {\em image} of a system in its heat bath, was earlier proposed by Papadodimas and Raju in~\cite{Papadodimas:2012aq}. However, their conclusions differ from ours.}

\section*{Thermo-field dynamics and black holes}

Ref.~\cite{doi:10.1142/S0217979296000817} developed the technique of thermo-field doubling in which a thermal state described by a density matrix on Hilbert space $\mathcal H$ can be ``doubled" to get a pure state -- the thermo-field vacuum
\be
\rho_\beta = \f{1}{Z} \sum_E e^{-\beta E} |E \rangle \langle E| ~~\rightarrow ~~|\Psi_\beta \rangle =\f{1}{\sqrt{Z}} \sum_E e^{-\beta E/2} |E \rangle \otimes | \tilde E \rangle
\ee
living on a doubled Hilbert space $\mathcal H \otimes  \mathcal {\tilde H}$.  Here $Z$ is the partition function. We adopt the notation of denoting the quantities in the first/ original system without a tilde and the ones in the second copy with tilde.
If one traces over the the second copy one recovers the thermal density matrix. All correlators calculated on the original copy can be recovered from the thermo-field vacuum
\be
Tr[ \rho_\beta  \hat O ] = \langle \Psi_\beta |  \hat O | \Psi_\beta \rangle.
\ee
It should be noted that in~\cite{doi:10.1142/S0217979296000817} it was emphasized that the second copy is to be understood as a fiction invented to make certain calculations simpler and has no physical meaning.

\begin{figure}[htbp]
\begin{center}
\includegraphics[scale=.4]{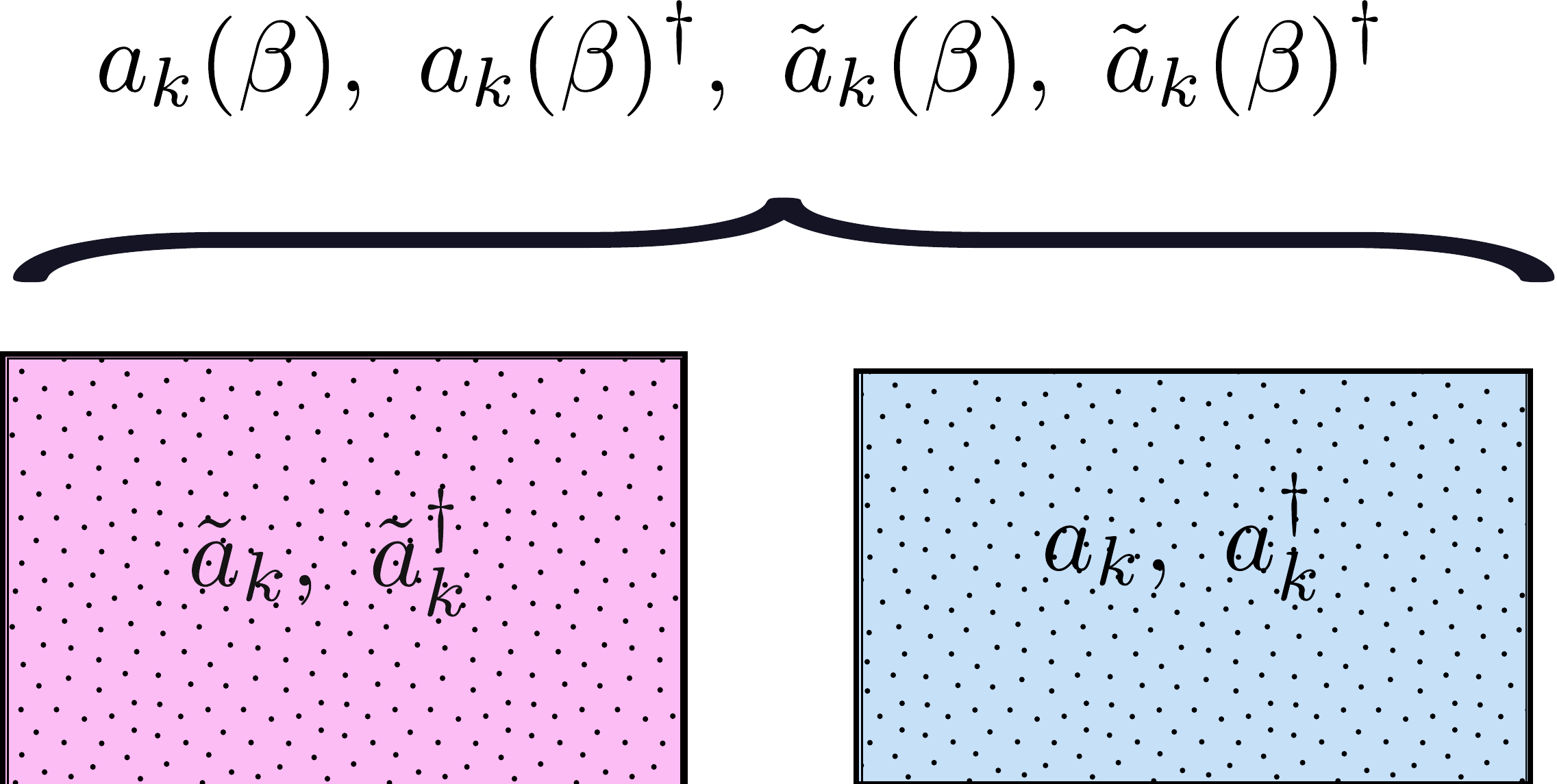}
\caption{The left box is the thermo-field double copy of a thermal state of the right box. The creation operators of the thermo-field vacuum have the interpretation of creating a particle in one box or annihilating one in the other on top of the thermal background. In the original thermo-field dynamics paper~\cite{doi:10.1142/S0217979296000817} anything not confined to the right box is a fiction.}
\label{FreeBosons}
\end{center}
\end{figure}

For our purposes it is instructive to consider thermo-field doubled system for a gas of free thermal bosons. The system is described by the usual creation and annihilation operators and a second set is introduced for the second copy. The thermo-field vacuum state is found to be annihilated by two sets of annihilation operators and the corresponding creation operators have the interpretation of creating particles in the one system or a hole in the other. This is shown in Figure~\ref{FreeBosons}. 

\begin{figure}[htbp]
\begin{center}
\subfigure[]{
\includegraphics[scale=.3]{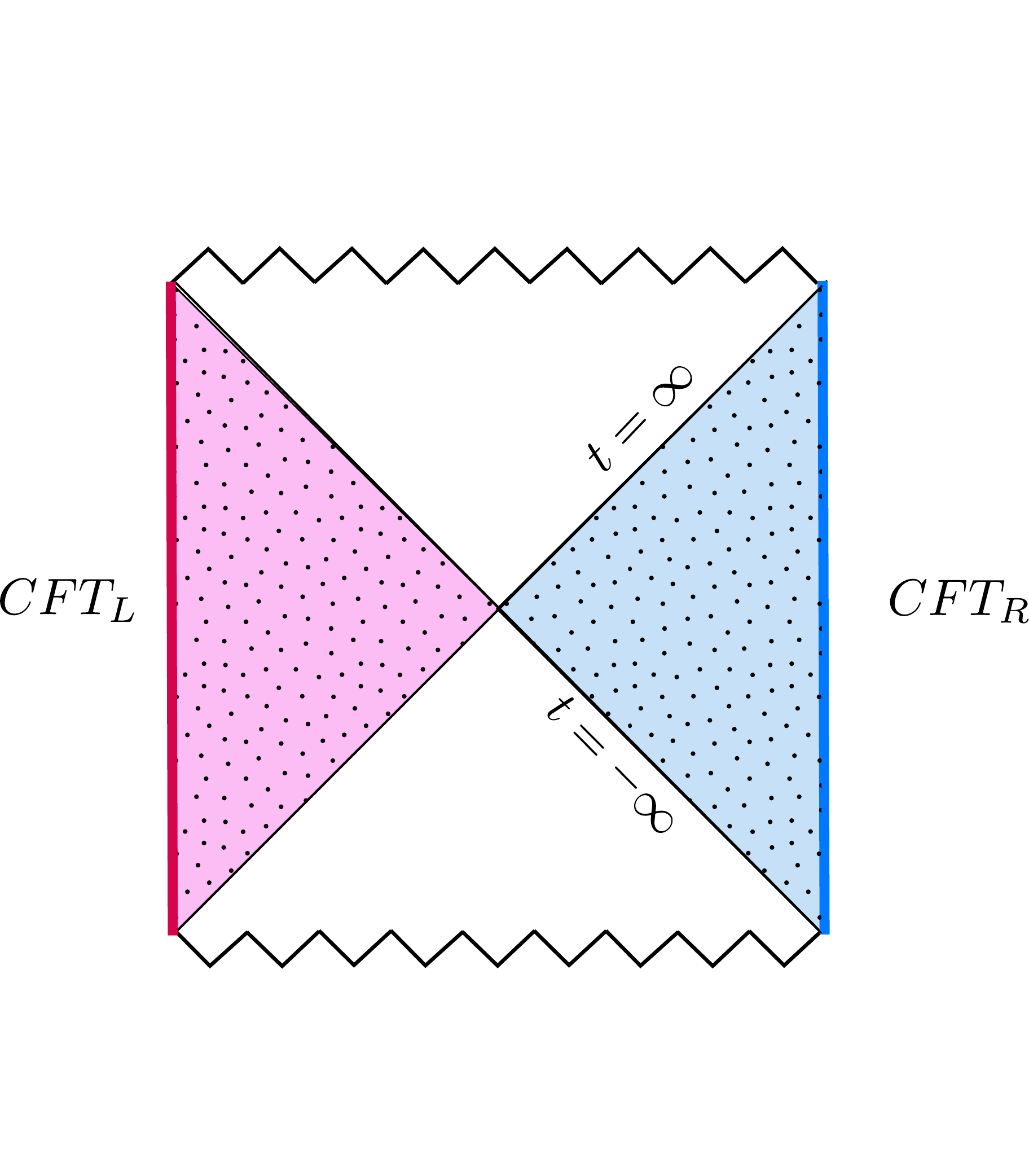}} \hspace{1in}
\subfigure[]{ 
\includegraphics[scale=.3]{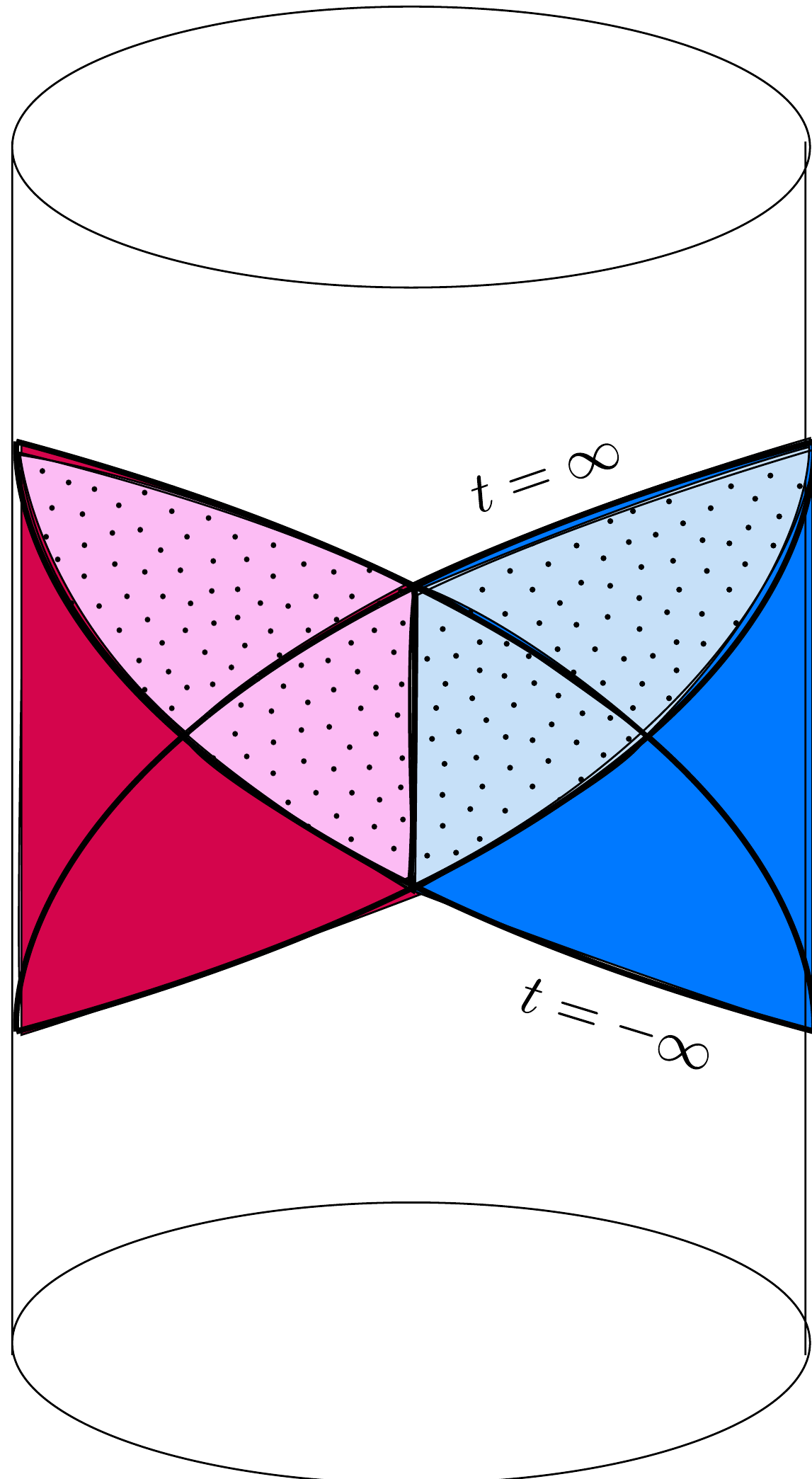} } 
\caption{In (a) the dual description of the eternal $AdS_{d+2}$ black hole involves two copies CFT on $S^d \times R$ which are in a thermo-field vacuum state. In (b) the dual of global $AdS_{d+2}$ may be described as the ground state of a CFT on $S^d \times R$ or alternatively as a thermo-field vacuum state on two CFTs on $H^d \times R$.}
\label{Horizons}
\end{center}
\end{figure}

In \cite{Israel:1976ur} Israel showed that as far as observers on one side of the horizon of a eternal Schwarzschild black hole are concerned, the other side behaves as a thermo-field doubled copy. The same is true for accelerating observers in Minkowski space from whose point of view the other side is the thermo-field doubled copy of their thermal Rindler state. Maldacena put Israel's ideas in the context of AdS/CFT in~\cite{Maldacena:2001kr} and proposed that the dual to an eternal $AdS_{d+2}$ black hole, which has two asymptotic boundaries, involves two CFTs on $S^d \times R$ which are in a thermo-field vacuum state. This is shown in Figure~\ref{Horizons}a. Further, it was shown in~\cite{Czech:2012be} that the dual to global $AdS_{d+2}$ can be described alternately as the ground state of the CFT on the $S^d \times R$ or as a thermo-field vacuum state on the two hyperbolic CFTs on $H^d \times R$. The duals to the individual CFTs are regions accessible to accelerating observers in the bulk.
 This is shown in Figure~\ref{Horizons}b.

\begin{figure}[htbp]
\begin{center}
\includegraphics[scale=.4]{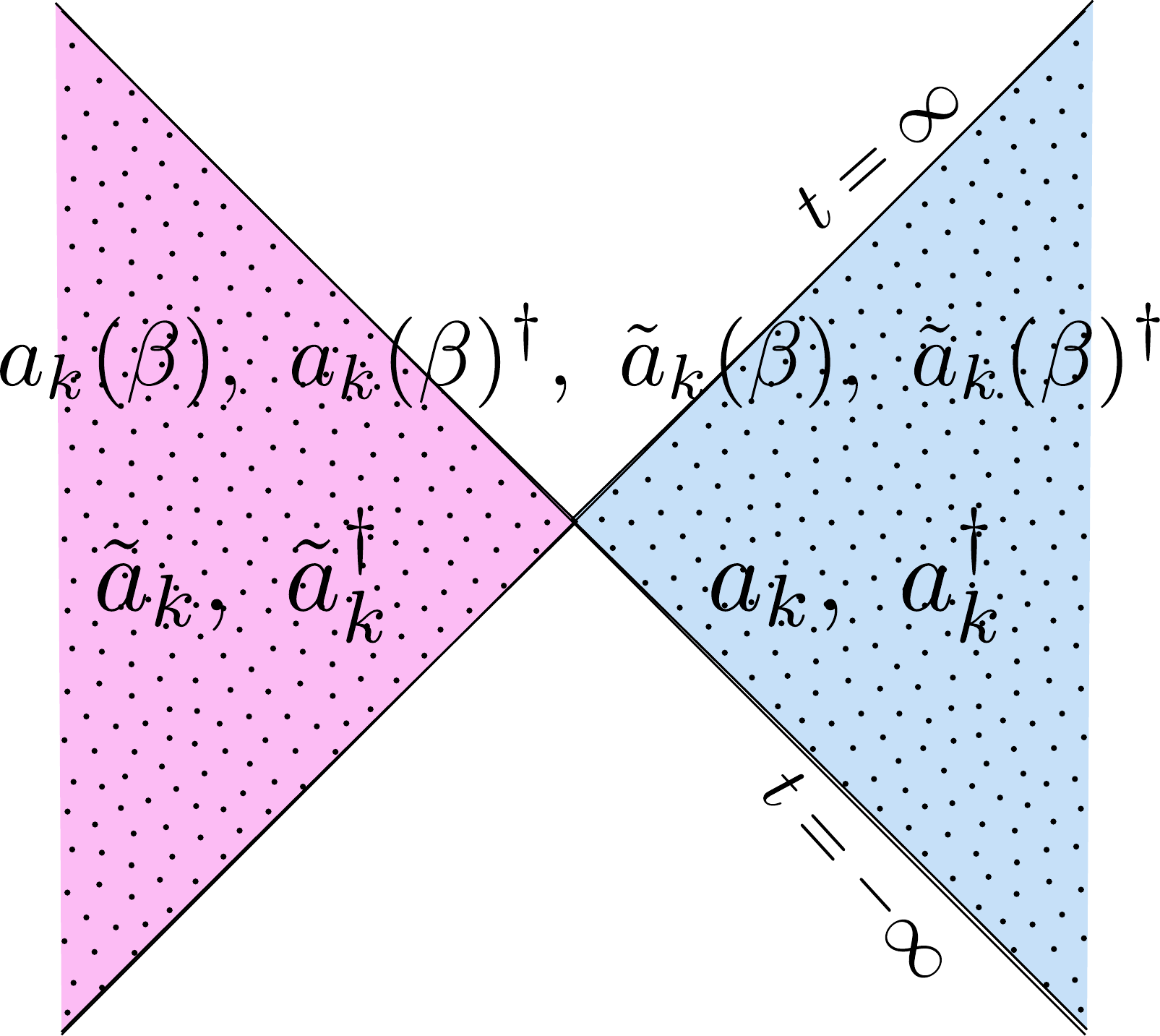}
\caption{The essential structure of the thermo-field dynamics interpretation of spacetimes with horizons. The two theories have their times reach $\pm \infty$ at the horizon. }
\label{FreeBosonsHorizons}
\end{center}
\end{figure}

The essential structure of the these examples is shown in Figure~\ref{FreeBosonsHorizons}.
Unlike the original thermo-field dynamics example shown in Figure~\ref{FreeBosons}, the other copy is  not fictitious anymore. To better see this note that observers created in the left and right wedges can pass through the future horizons and meet in the forward wedge. There is no analogue of the forward wedge in Figure~\ref{FreeBosons}. The creation operators $a_k^\dagger(\beta)$ and $\tilde a_k^\dagger(\beta)$ create particles spanning space-like slices in the left and right wedge which can be evolved into the forward and backward wedges. Thus, the thermo-field doubling is not just a fictitious tool in geometries with horizons.

\section*{Black holes or firewalls?}

The CFTs in Figures~\ref{Horizons}a and~\ref{Horizons}b are understood to be non-interacting as they are outside each others light cones. Note that while  the CFTs in Figure~\ref{Horizons}b  live on the same sphere, by being defined in finite global time, they are nevertheless non-interacting. However, from the dual description (Figure~\ref{FreeBosonsHorizons}) it is clear that an excitation which crosses the future horizon requires both the CFTs for its description. That the CFT time for crossing is infinite is related to the Schwarzschild coordinates not covering the full manifold and do not stop an infalling observer from crossing the horizon. It is this ability of an excitation with initial support largely on one CFT to leak behind the horizon and require support on the other CFT which makes thermo-field doubles non-fictitious. 

\begin{figure}[htbp]
\begin{center}
\subfigure[]{
\includegraphics[scale=.15]{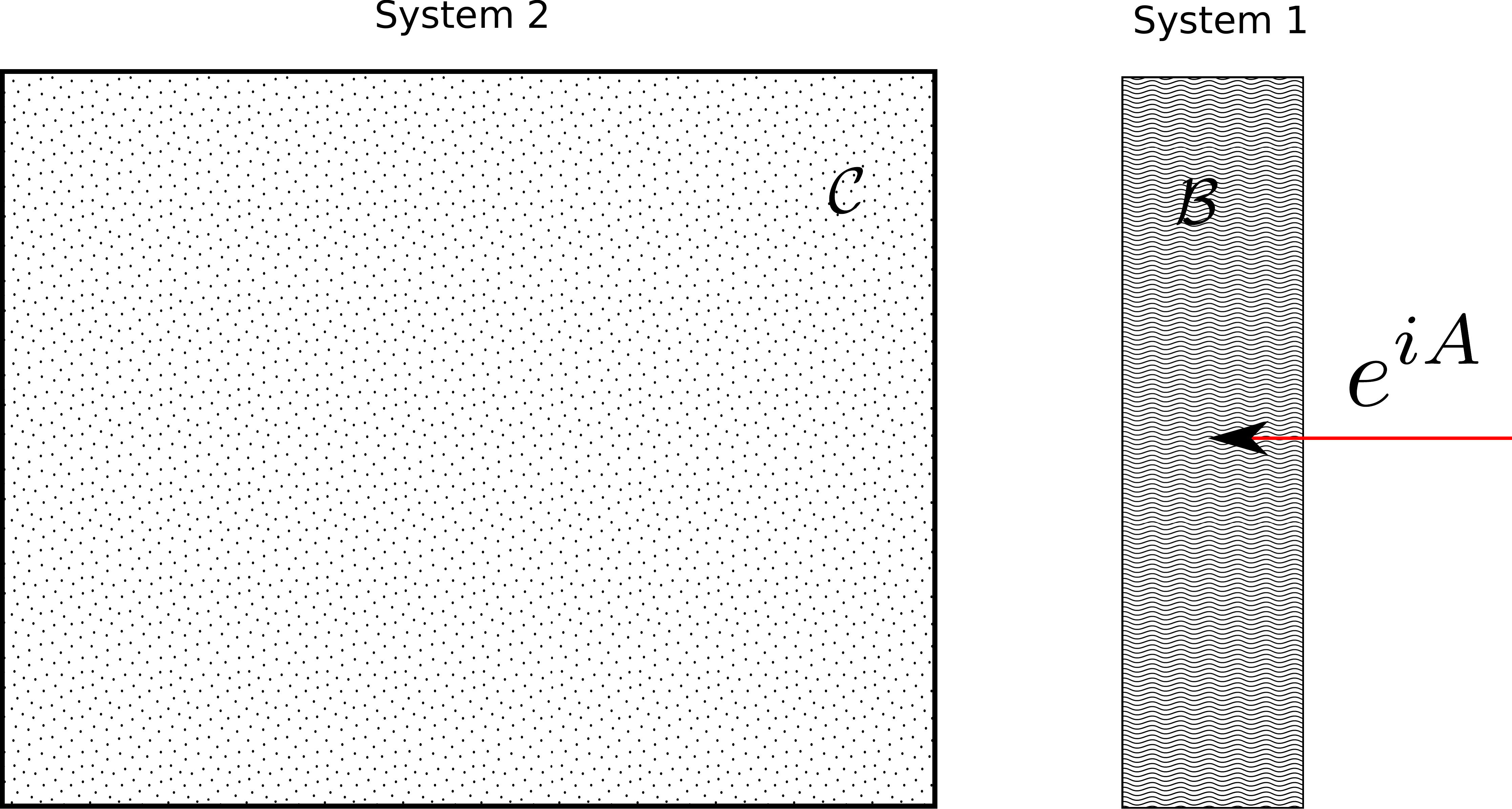}}
\subfigure[]{ 
\includegraphics[scale=.15]{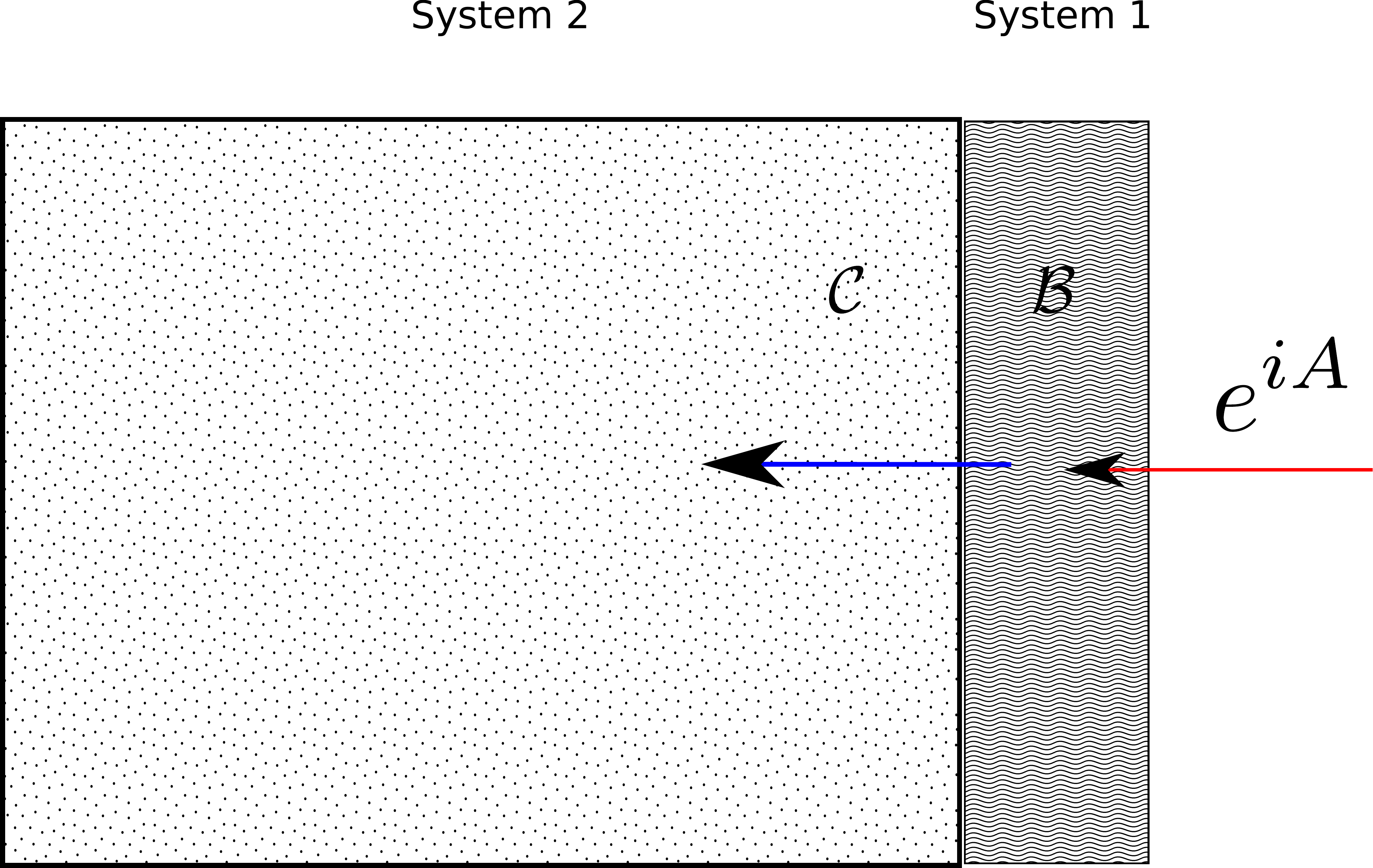} } 
\caption{A thermal system may be (a) weakly coupled to an arbitrary heat bath or (b) strongly coupled to an efficient {\em fast scrambling} heat bath.  Evolution of an excitation on the system will evolve differently in these two cases.}
\label{KRSplit}
\end{center}
\end{figure}

A system becomes thermal by coupling to a heat-bath and the combined state of the system and its heat-bath is
\be
|\Psi \rangle = \sum_E e^{-\beta E/2} |E \rangle \otimes |E\rangle_{\text{HB}}. 
\ee
in the Schmidt basis. The thermo-field double may be thought to be the ``scrambled image" of the system in the heat bath. For a weak interaction between the system and its heat bath, this identification has no consequences as the evolution of the excitation is described to a very good approximation by the system alone as shown in Figure~\ref{KRSplit}a. However, if the system and heat-bath are strongly coupled then an excitation on the system will leak into the heat bath very quickly as shown in \ref{KRSplit}b. Then the heat-bath will play a role in describing the evolution of the excitation.

There are reasons to believe that the CFT dual to a gravity theory is strongly coupled \cite{Aharony:1999ti}. If we assume that the heat-bath is also a strongly coupled CFT and is strongly coupled to the CFT system under consideration then Figures~\ref{Horizons}a and~\ref{Horizons}b seem to be a realization of the situation considered in Figure~\ref{KRSplit}a. The gravitational picture seems to suggest that the doubled copy/ image plays a role {\em before} the rest of the heat bath does. 

Note that in Figure~\ref{Horizons}b the double copy is the entire heat bath. The lack of a bigger heat bath seems to be correlated with the lack of a singularity and is suggestive that the singularity represents an excitation evolving to the point that it requires more degrees of freedom than just the system and its doubled copy/ image can provide.

It stands to reason that if a strong coupling between a CFT and its heat bath, which also has to be a CFT, is required for smoothness across horizons then a generic heat bath will result in  a firewall being the dual to a thermal CFT.

\section*{Conclusion}

Geometries with horizons are naturally described in terms of the thermo-field double formalism. However, the thermo-field doubled system is not fictitious in these cases and the entanglement between the system and its double are necessary to describe physics in the future of the horizon. The thermo-field double should be viewed as the ``image" of the system in the heat bath and smoothness across the horizon or the presence of firewalls is governed by coupling between the system and the heat bath and the properties of the heat bath.

\section*{Acknowledgements}

I would like to thank Steve Avery, Bartek Czech, Kyriakos Papadodimas and Erik Verlinde for useful discussions. My work is supported by the ERC Advanced Grant 268088-EMERGRAV.

\bibliographystyle{toine}
\bibliography{Papers}

\end{document}